\documentclass[preprint,preprintnumbers,amsmath,amssymb,prl,superscriptaddress]{revtex4-1}

\usepackage{physics}
\usepackage{graphicx}
\usepackage{dcolumn}
\usepackage{bm}
\usepackage{epstopdf}
\usepackage[colorinlistoftodos]{todonotes}
\usepackage{float}
\usepackage[T1]{fontenc}
\usepackage[utf8]{inputenc}
\usepackage{parskip}
\usepackage{physics}
\usepackage{mathtools}

\newcommand*\chem[1]{\ensuremath{\mathrm{#1}}}

\begin{document}


\title{Hard superconducting gap and vortex-state spectroscopy in NbSe$_2$ van der Waals tunnel junctions}

\author{T. Dvir}
\affiliation{The Racah Institute of Physics, the Hebrew University of Jerusalem, Israel}

\author{F. Massee}
\affiliation{Laboratoire de Physique des Solides (CNRS UMR 8502), Bâtiment 510, Université Paris-Sud/Université Paris-Saclay, 91405 Orsay, France}

\author{L. Attias}
\affiliation{The Racah Institute of Physics, the Hebrew University of Jerusalem, Israel}

\author{M. Khodas}
\affiliation{The Racah Institute of Physics, the Hebrew University of Jerusalem, Israel}

\author{M. Aprili}
\affiliation{Laboratoire de Physique des Solides (CNRS UMR 8502), Bâtiment 510, Université Paris-Sud/Université Paris-Saclay, 91405 Orsay, France}

\author{C. H. L. Quay}
\affiliation{Laboratoire de Physique des Solides (CNRS UMR 8502), Bâtiment 510, Université Paris-Sud/Université Paris-Saclay, 91405 Orsay, France}

\author{H. Steinberg}
\affiliation{The Racah Institute of Physics, the Hebrew University of Jerusalem, Israel}

\date{\today}

\maketitle

\textbf{Device-based tunnel spectroscopy of superconductors was first performed by Giaever, whose seminal work provided clear evidence for the spectral gap in the density of states (DOS) predicted by the Bardeen-Cooper-Schrieffer (BCS) theory~\cite{Giaever1960a}. Since then, tunnel-barrier-based heterostructure devices have revealed myriad physical phenomena~\cite{McMillan1968,Rowell1973,Dynes1978,Dynes1984TunnelingTransition,Fert,Kastner,Wolf} and found a range of applications~\cite{Giazotto_Rev,vanderWal,Devoret}. Most of these devices rely on a limited number of oxides, which form high-quality insulating, non-magnetic barriers.
These barriers, however, do not grow well on all surfaces. Promising alternatives are van der Waals (vdW) materials~\cite{Geim2013}, ultrathin layers of which can be precisely positioned on many surfaces~\cite{Dean2010}; they have been shown to form tunnel barriers when engaged with graphene~\cite{Amet2012,Britnell2012,Chandni2016}.
Here we demonstrate that vdW semiconductors MoS$_2$ and WSe$_2$ deposited on the superconductor NbSe$_2$ form high quality tunnel barriers, with transparencies in the $10^{-8}$ range. Our measurements of the NbSe$_2$ DOS at 70mK show a hard superconducting gap, and a quasiparticle peak structure with clear evidence of contributions from two bands~\cite{Schopohi1977a,McMillan1968,Noat2015}, with intrinsic superconductivity in both bands. In both perpendicular and parallel magnetic fields, we observe a sub-gap DOS associated with vortex bound states~\cite{Nakai2004b, Nakai2006a}. The linear dependence of the zero-bias signal on perpendicular field allows us to confirm the $s$-wave nature of superconductivity in NbSe$_2$. As vdW tunnel barriers can be deployed on many solid surfaces, they extend the range of superconducting and other materials addressable not only by high resolution tunneling spectroscopy but also non-equilibrium and/or non-local transport~\cite{Clarke,Jedema,Quay,Hubler}.}

Conductance-voltage characteristics obtained when tunnelling across normal metal-insulator-superconductor (NIS) junctions (as in the Giaever experiment) are dominated by strong quasiparticle peaks at energies corresponding to $\pm\Delta$, where $\Delta$ is the superconducting gap. Below the gap, in BCS superconductors, the conductance signal due to quasiparticles is strongly suppressed. Conductance at these energies might be due to finite quasiparticle lifetimes in materials with strong electron-phonon coupling~\cite{Dynes1978} or sub-gap quantum states, e.g. Caroli-de Gennes-Matricon vortex bound states in Type II superconductors~\cite{Caroli1964}. Alternatively, in systems in which superconductivity is induced by proximity, subgap spectroscopy has revealed Andreev Bound States~\cite{Andreev1964TheSuperconductors,deGennes1963ElementaryContact,Rowell1973, Pillet2010, Dirks2011}. Recent interest in tunneling at sub-gap energies has been driven by the search for Majorana and other exotic states in proximitised topological insulators~\cite{Kane2008}, graphene~\cite{Kopnin,Khaymovich2009b} and semiconductor nanowires~\cite{Das2012, Mourik2012}.

Such experiments are critically dependent on the ability to resolve spectral features above the sub-gap background signal. Sub-gap tunneling across NIS junctions with transparent barriers can arise due to two-electron~\cite{Schrieffer1964TheorySuperconductivity} or Andreev~\cite{Blonder1982} processes. In more opaque junctions it is was often associated with barrier defects~\cite{Kleinsasser1993a}, although more recent work points to diffusive Andreev processes~\cite{Greibe2011} and environment-assisted tunneling \cite{Pekola2010Environment-AssistedStates,DiMarco2013}. $G_0 R_N$, the zero-energy conductance times the normal state resistance is a useful figure of merit, and has been reported to be $\approx 1/100,000$ in junctions based on \chem{Al_2O_3}~\cite{Greibe2011}; however, reaching hard-gap junctions in other systems has proven to be challenging. In semiconducting nanowires, for example, only the recent development of epitaxial barriers resulted in strongly suppressed sub-gap signal~\cite{Chang2015}.

\begin{figure}
	\centering
	\includegraphics[width=0.6\textwidth]{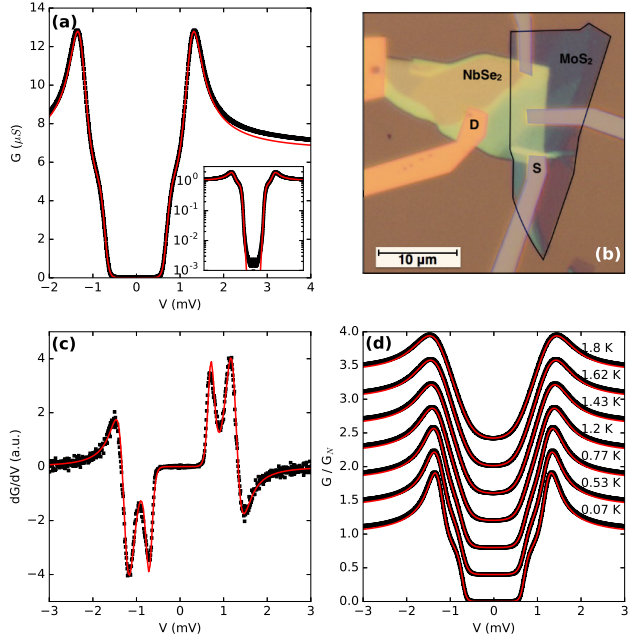}
	\caption[caption]{\textbf{Differential conductance of a NIS tunnel junction. a,} $dI/dV$ vs. $V$ as measured on the device shown in \textbf{b} (black) and a fit to the SSM model (red, see details in the text). Inset: $dI/dV$ on a logarithmic scale. \textbf{b,} Optical image of the tunnel junction device. The yellow-green flake is a 50-20 nm thick \chem{NbSe_2}  (20 nm at the source electrode) and the purple-blue flake is a 4-5 layer \chem{MoS_2} (marked by black frame). Au electrodes are deposited on the left to serve as ohmic contacts (yellow) and on the right to serve as tunnel electrodes (purple). \textbf{c,} $d^2I/dV^2$ of panel (a), and the fit to the SSM model.  \textbf{d,} $dI/dV$ curves taken at different temperatures (black) and fits to the model (red), vertically shifted for clarity.}
	\label{figure1}
\end{figure}

Using vdW layered materials as tunnel barriers greatly expands the range of addressable materials, in particular to those not easily covered by oxides~\cite{Amet2012,Britnell2012,Chandni2016}. These barriers can be deposited using the “dry transfer” fabrication technique, which allows successive stacking of multiple flakes of vdW materials to form heterostructures~\cite{Geim2013,Dean2010}. Our devices are NIS tunnel junctions with either \chem{MoS_2} or \chem{WSe_2} -- both vdW materials -- as the insulating barrier. The barrier material is placed on top of \chem{2H-NbSe_2} (hereafter \chem{NbSe_2}), a vdW superconductor with $T_c\approx 7.2$ K. This insulator-superconductor structure is contacted by Au electrodes, which either directly engage the \chem{NbSe_2} flake to create ohmic contacts; or else are deposited over the barrier (Fig. \ref{figure1}b), forming the N of the NIS junction. A voltage $V$ is applied across the junction and the current $I$ across it measured.

Fig. \ref{figure1}b shows a typical junction (`Device A') consisting of a 4-5 layer thick \chem{MoS_2} barrier (Supplementary Section 2) with a transparency $\mathcal{T}\sim10^{-8}$ (Supplementary Secion 6). Its normal state conductance $G_N=$ 7$\mu$S for an area $A =$ 1.6 $\mu m^2$.  Fig. \ref{figure1}a. shows the differential conductance $G = dI/dV$ as a function of $V$ (normalized to $G_N$) obtained with Device A, at $T = 70 mK$. This spectrum has two striking features: First, the very low sub-gap conductance ($G_0 R_N \approx 1/500$), evident in the logarithmic plot presented at the inset. The residual conductance is likely due to environment-assisted tunneling (Supplementary Section 7). Second, the intricate structure of the quasiparticle peak. This spectrum differs from a standard BCS DOS by having a relatively low quasiparticle peak and a shoulder at lower energies. The latter feature can be clearly seen in the second derivative (Fig. \ref{figure1}c) where the slope separates into a double peak feature. In what follows, we begin by analyzing the structure of the quasiparticle peak using the two-band model. We then present measurements in magnetic field, where the low sub-gap background allows us to observe vortex bound states.

Two-band superconductivity was first discussed theoretically by Suhl et al.~\cite{Suhl1959}, who considered distinct BCS coupling strengths for each band $i = 1, 2$ and allowed for Cooper-pair tunneling between bands. Schopohl and Scharnberg expanded on this, including inter-band single-electron scattering (parametrised by $\Gamma_{12},\Gamma_{21}$), in addition to Cooper pair tunneling~\cite{Schopohi1977a}. These interband processes give rise to modified pairing amplitudes $\Delta_{i}^0$, resulting in a model corresponding to McMillan's description of the proximity effect between a superconductor and a normal metal~\cite{McMillan1968}. The resulting ``SSM'' model has successfully been used to fit tunneling conductance data from \chem{MgB_2} SIS junctions~\cite{Schmidt}, as well as scanning tunneling spectroscopy data from \chem{NbSe_2}, indicating the two-band nature of these materials. 

In the SSM model, the superconducting gaps $\Delta_i (E)$ in the two bands $i$ are found by solving the coupled equations

\begin{eqnarray}
\Delta_i (E) = \frac{\Delta_i^0 + \Gamma_{ij} \Delta_j(E)/\sqrt{\Delta_j^2(E)-E^2}}{1 + \Gamma_{ij} /\sqrt{\Delta_j^2(E)-E^2}}
\label{selfcons}
\end{eqnarray} 
whereas the DOS of each band is given by
\begin{eqnarray}
N_S^i(E)  = N_i(E_F) \Re{\frac{|E|}{\sqrt{\Delta_i^2(E)-E^2}}}.
\end{eqnarray}

Here $N_i(E_F)$ is the DOS at the Fermi energy in the normal state in band $i$. The conductance is calculated by convolution of the DOS with the derivative of a Fermi-Dirac distribution with temperature $T$, accounting for a ratio $\eta$ between the tunneling matrix elements for the two bands,

The best fit to our data with the above equations is shown in Fig. \ref{figure1}a, where the following fitting parameters are extracted :  $\Delta_1^0 = 1.23\pm0.01\  \mathrm{meV}$, $\Delta_2^0 = 0.36\pm0.05\  \mathrm{meV}$, $\Gamma_{12} = 1.1 \pm 0.2\ \mathrm{meV}$, $\Gamma_{21} = 0.38\pm0.07\ \mathrm{meV}$,  $T = 0.52\pm 0.05 \ \mathrm{K}$, and $\eta = 1:0.13$. As seen in the figure, our fit is remarkably precise -  successfully reproducing both the first and second derivative experimental curves. It allows us to confirm the SSM model and determine the various parameters with unprecedented fidelity. The most salient feature in our fit is the identification of intrinsic superconducting pairing in the second band, manifest as $\Delta_2 > 0$. This yields a better fit to the data (Supplementary Fig. 3)
than the alternative ($\Delta_2 = 0$), which corresponds to induced pairing~\cite{Noat2015}. These same parameters yield successful fits also at elevated temperatures, while changing only $T$ (panel (d)). At the lowest temperature, however, the fitting temperature exceeds the expected electron temperature. It is unlikely that this is due to junction heating, and we suggest it is associated with inhomogeneity in $\Delta$ (Supplementary Section 8).

\begin{figure} 
\centering
\includegraphics[width = 0.6\textwidth]{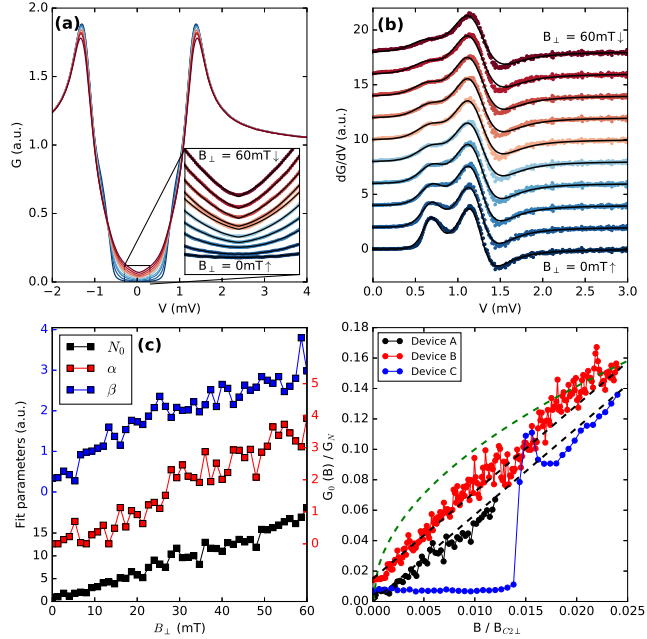}
\caption{\textbf{Response of the tunneling conductance to perpendicular magnetic fields. 
a,}   $dI/dV$ curves at increasing magnetic field $B_\perp$ perpendicular to the \chem{NbSe_2} layers. Inset:  magnification of the sub-gap tunneling spectrum, fit to a quadratic model $N_0 +\alpha |V|+\beta V^2$. 
\textbf{b,} $d^2I/dV^2$ of data in (a), fit using the 2-band SSM model. 
\textbf{c,} $B_\perp$-dependence of the quadratic fit parameters.
\textbf{d,} Normalized zero-bias conductance $G_0(B)/G_N$ vs. normalized field $B/ B_{c2}$. The device discussed here (Device A, black) is compared to two other devices (B,C), all showing linear dependence with the same slope.  This data clearly fails to fit a square root dependence (green, dashed line).
}
 \label{Figure2}
\end{figure}

We now turn to the response of the tunneling spectrum to perpendicular magnetic field $B_\perp$, shown in Fig.~ \ref{Figure2}.  Panel (a)  shows a collection of $dI/dV$ curves taken at perpendicular magnetic fields $0 \leq B_\perp \leq 60 $ mT. The data show that $B_\perp$ has two observable consequences: (i) it suppresses the lower energy shoulder of the quasiparticle peaks, seen most clearly in the $d^2I/dV^2$ plots in panel (b), and (ii) it increases the subgap signal (for $|V|<0.5\ \textrm{mV}$), seen in the inset.  We find that  in this low magnetic field it is possible to fit the modified spectra using the same 2-band model as the zero field data. The fit is superimposed on the data in panel (b), and details of the fitting parameters are discussed in Supplementary Section 5.

In type-II superconductors above  $B_{c1}$ vortices penetrate the sample. In this so-called ``mixed state'' the superconductor consists of quasi-normal vortex cores, and a gapped inter-vortex area. Due to the hard gap, our measurement spectrally differentiates between these regions: at low bias, sub-gap tunneling takes place only near the vortex cores.  At higher bias, quasiparticle tunneling occurs at the entire area of the sample. In the remainder of this letter we provide further evidence that the sub-gap signal is associated with vortex-bound states. 

Close inspection of the low bias region in Fig.~\ref{Figure2}a (inset) shows the onset of V-shaped spectra at low magnetic fields. As shown by Nakai et al.~\cite{Nakai2006a} such spectra are inherent to the integrated spectral weight of vortex-bound states, 
regardless of the symmetry of the order parameter. These comprise of zero-bias spectral weight $N_0$ and annular states centered at an energy-linear radius. Upon polar integration, the latter yield the term $\alpha |V|$.  

Finally, ref.~\cite{Nakai2006a} also identifies a quadratic term, $\beta V^2$. We carry out the same fit:  $G(V) = N_0 + \alpha |V| + \beta V^2$, and extract the dependence of $N_0$, $\alpha$ and $\beta$ on $B_\perp$. The $B_\perp$-dependent fitting parameters are shown in Fig.~\ref{Figure2}c, where it is evident that all three increase monotonously with $B_\perp$. We interpret $N_0(B_\perp)$ as the product of the zero-energy DOS at each vortex, times the number of vortices accessible to the tunnel junction. For $s$-wave superconductors $N_0(B_\perp)$ is linear in field~\cite{Nakai2004b}, and can gauge the number of vortices in the junction. The dependence of $\alpha(B_\perp)$, which also exhibits a linear increase with $B_\perp$, can be interpreted in a similar way, since it represents a population of off-center states which are associated with individual vortices. The interpretation of $\beta(B_\perp)$ is somewhat less straightforward. We will argue below that it is associated with currents induced around the vortices. We conclude that the subgap signal, and especially the linear term in the signal, is a good proxy for probing vortex penetration of the sample.  

\begin{figure}\includegraphics[width =0.6\textwidth]{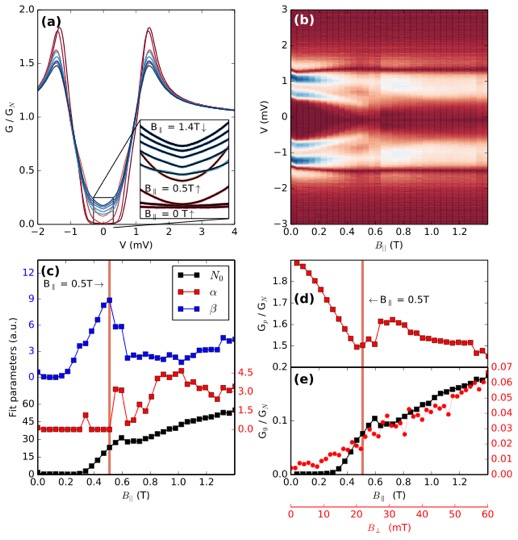}
\caption{\textbf{Response of the tunneling conductance to parallel magnetic fields. 
a,} $dI/dV$ curves at increasing magnetic field parallel to the \chem{NbSe_2} layers ($B_\parallel$). Inset: magnification of the sub-gap tunneling spectrum, fit to a quadratic model $N_0 +\alpha |V|+\beta V^2$. 
\textbf{b,} Color map of $d^2I/dV^2$ vs.  $V$ and $B_\parallel$. 
\textbf{c,} $B_\parallel$-dependence of the quadratic fit parameters. The transition field, $B_\parallel = 0.5$ T, is marked in light red.
\textbf{d,} Quasiparticle peak height $G_p$ vs. $B_\parallel$.
\textbf{e,} Zero bias spectral weight vs. $B_\parallel$. The transition field, $B_\parallel = 0.5$ T, is marked in light red. This is compared to the zero bias spectral weight Vs. $B_\perp$ (red).
}
\label{Figure3}
\end{figure}

We repeat the same measurement and analysis for magnetic field applied parallel to the sample, $B_\parallel$ (Fig.~ \ref{Figure3}), up to 1.5T. Once again, we can follow the evolution of the low energy shoulder, which appears as a peak in $d^2I/dV^2$ (panel b). Here, unlike the $B_\perp$ case, this feature remains resolved as $B_\parallel$ increases, but shifts to lower energies. We attribute this to Abrikosov-Gor'kov depairing~\cite{AbrikosovA.A.1961ContributionImpurities,Maki1964TheCurrents, Levine1967DensityTunneling, Millstein1967TunnelingField, Anthore2003}, though here the effect is somewhat more complex than what was seen in previous works due to the 2-band nature of \chem{NbSe_2}~\cite{Kaiser1970McMillanAlloys}.

The signal changes abruptly at $B_\parallel=0.5 T$, which we associate with $B_{c1}^\parallel$. It is manifest as an increase of the height of the quasiparticle peak (panel d) and a sharp increase of the zero bias signal (panels c,e). We rule out the possible contribution of residual perpendicular fields, due to misalignment of the sample with the parallel field; this is compensated to better than 0.5\% of the parallel field. Tunneling investigation away from perpendicular fields was carried out by Hess et al.~\cite{Hess1994}, who observed complex flux lattices at various angles, and strictly parallel flux lines were observed in ref.~\cite{Fridman2011a}, where the Meissner currents indicated the positions of buried vortices. However, all these studies utilized bulk samples, whereas our sample thickness is $d = 20 nm \ll \lambda_L$, imposing spacial restrictions on the Meissner currents.

To probe the nature of this regime, we apply the same sub-gap analysis carried out for $B_\perp$ (Inset to Fig.~ \ref{Figure3}a). For $B_\parallel < B_{c1}^\parallel$, the signal is described by a parabola with zero-offest, such that both $N_0$ and $\alpha$ remain small, while $\beta$ increases. For $B_\parallel > B_{c1}^\parallel$, we find that $\beta$ drops sharply. This is accompanied by an increase in $N_0$ and $\alpha$, consistent with the onset of vortex tunneling. The drop in $\beta$ suggests that the parabolic term is partly a consequence of the Meissner currents, which drop sharply above $B_{c1}$. The appearance of vortex-bound $N_0$ and $\alpha$ terms suggests that vortex sub-gap tunneling is taking place, similar to the $B_\perp$ case. This in turn, indicates tunneling accessibility to points where flux lines enter and exit the sample, likely due to defects or variations in thickness. In the alternative scenario, where flux lines are strictly parallel and are buried under the surface~\cite{Fridman2011a}, vortex-bound states would not be observable.

We now turn to discussion of the zero-bias conductance, and its dependence on $B$. For $B_\perp$ (Fig.~\ref{Figure2}d), it is likely that the onset of vortex penetration is very close to $B=0$. $N_0(B_\perp)=G_0(B_\perp)$ increases linearly with $B_\perp$, consistent with a constant increase of vortex population. Following ref.~\cite{Nakai2004b}, we present $G_0(B)/G_N$ vs. $B/B_{c2}$ (dimensionless units). The data of Device A (black dots), follow a slope $dN_0(B_\perp)/dB_\perp \approx 6$,  reflecting a rapid increase in bound-state spectral weight. This slope appears to be generic, as two other devices (B and C, red and blue dots) exhibit a similar slope, noting that in Device C there is a finite onset field. For $s$-wave superconductors one expects minimal vortex overlap, resulting in a minimal slope of 1.2~\cite{Nakai2004b}, where exceeding this value could indicate deviations from perfect isotropy. These, however, would modify the quasi-particle peak structure. Based on the broadening we actually observe, the anisotropy remains capped by 1.2 (Supplementary Section 8), and hence cannot be the origin of the high spectral weight we observe. It is also not compatible with line-node anisotropy since the zero bias signal clearly deviates from a square root dependence expected in this case.  A possible explanation is a renormalisation of the local magnetic field at the junction due to flux focusing.

Our work opens up the possibility of using vdW barriers to investigate the density of states of other vdW materials, and in particular superconductors~\cite{Lu_MoS2_2015, Saito_MoS2_2015,Xi_2016}. As vdW tunnel barriers adhere readily to clean, flat surfaces, they could also be deposited on non-vdW (super)conductors. Furthermore, as the dry transfer technique does not involve solvents, and as the resulting device size is compatible with custom mechanical masks (thus eliminating the need for lithography), vdW tunnel barriers could perhaps also be deposited on organic (super)conductors and other fragile systems which have hitherto not been investigated in tunnel spectroscopy. Finally, we note that fabricating multiple, closely-spaced tunnel electrodes on the same device --- a feasible extension of our present methods --- will allow the investigation of many new systems under non-equilibrium conditions~\cite{Clarke,Gray,Jedema,Quay,Hubler}

\section*{Methods}
The vdW tunnel junctions were fabricated using the dry transfer technique \cite{Castellanos-Gomez2014}, carried out in a glove-box (nitrogen atmosphere). \chem{NbSe_2} flakes were cleaved using the scotch tape method, peeled on commercially available Gelfilm from Gelpack, and subsequently transferred to a \chem{SiO_2} substrate. \chem{MoS_2} and \chem{WSe_2} flakes were peeled in a same way, where thin flakes suitable for the formation of tunnel barriers were selected based on optical transparency. The barrier flake was then transferred and positioned on top of the \chem{NbSe_2} flake at room temperature. Ti/Au contacts and tunnel electrodes were fabricated using standard e-beam techniques. Prior to the evaporation of the ohmic contacts the sample was ion milled for 15 seconds. No such treatment was done with the evaporation of the tunnel electrodes. All transport measurements were done in a $^3$He–$^4$He dilution refrigerator with a base temperature of 70 mK. The AC excitation voltage was modulated at 17 Hz; its amplitude was 15$\mu$V at all temperatures. Measurement circuit details are provided in Supplementary Section 1.

\section*{Acknowledgements}
We thank P. Février and J. Gabelli for helpful discussions on tunnel barriers, and T. Cren for the same on \chem{NbSe_2}. This work was funded by a Maimonïdes-Israel grant from the Israeli-French High Council for Scientific \& Technological Research;  ERC-2014-STG Grant No. 637298. (TUNNEL); and an ANR JCJC grant (SPINOES) from the French Agence Nationale de Recherche. T.D. is grateful to the Azrieli Foundation for an Azrieli Fellowship. L.A. and M.K. are supported by the Israeli Science Foundation, Grant No. 1287/15.

\section*{Author contributions}
T.D. fabricated the devices. T.D. and C.Q.H.L. performed the measurements. All the authors contributed to data analysis and the writing of the manuscript.

\section*{Competing financial interests}
The authors declare no competing financial interests.

\bibliographystyle{naturemag}
\bibliography{Corrected_bib}

\pagebreak
\widetext
\begin{center}
\textbf{\large Supplemental Materials: Hard superconducting gap and vortex-state spectroscopy in NbSe$_2$ van der Waals tunnel junctions}
\end{center}
\setcounter{equation}{0}
\setcounter{figure}{0}
\setcounter{table}{0}
\setcounter{page}{1}
\makeatletter
\renewcommand{\theequation}{S\arabic{equation}}
\renewcommand\thesection{S\arabic{section}}
\renewcommand\thefigure{\textbf{S\arabic{figure}}}   
\renewcommand{\figurename}{\textbf{Supplementary Figure}}
\setcounter{secnumdepth}{1}

\section{Details of the measurement setup}

\begin{figure}[h]
	\centering
	\includegraphics[width=0.5\textwidth]{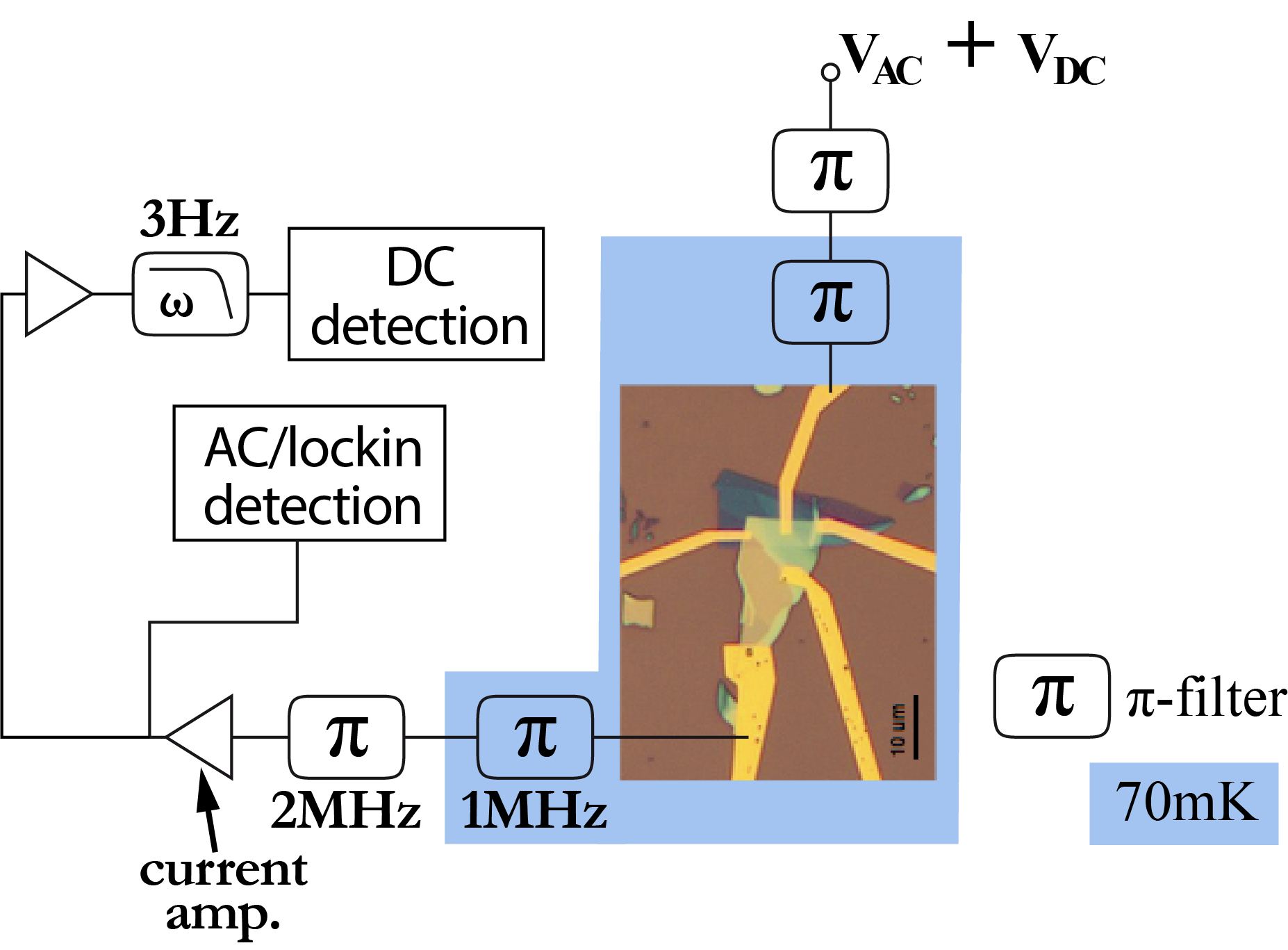}
	\caption[caption]{Detailed diagram of the measurement circuit used in the experiment.}
	\label{circuit}
\end{figure}

Figure~\ref{circuit} shows our measurement circuit in greater detail than was presented in the main text. All $\pi$-filters at low temperature have cutoff frequencies of 1MHz while those at room temperature have cutoff frequencies of 2MHz. The amplitude of the AC excitation $V_{AC}$ is 15$\mu$V in all the figures of the main text. Measurements at lower $V_{AC}$ showed that, between 2$\mu$V and 15$\mu$V, there was no discernible distortion of $G(V)$; the higher excitation voltage was thus chosen in order to have a better signal-to-noise ratio.

\section{Thickness and structure of the tunnel barrier}

\begin{figure}[h]
	\centering
	\includegraphics[width=0.5\textwidth]{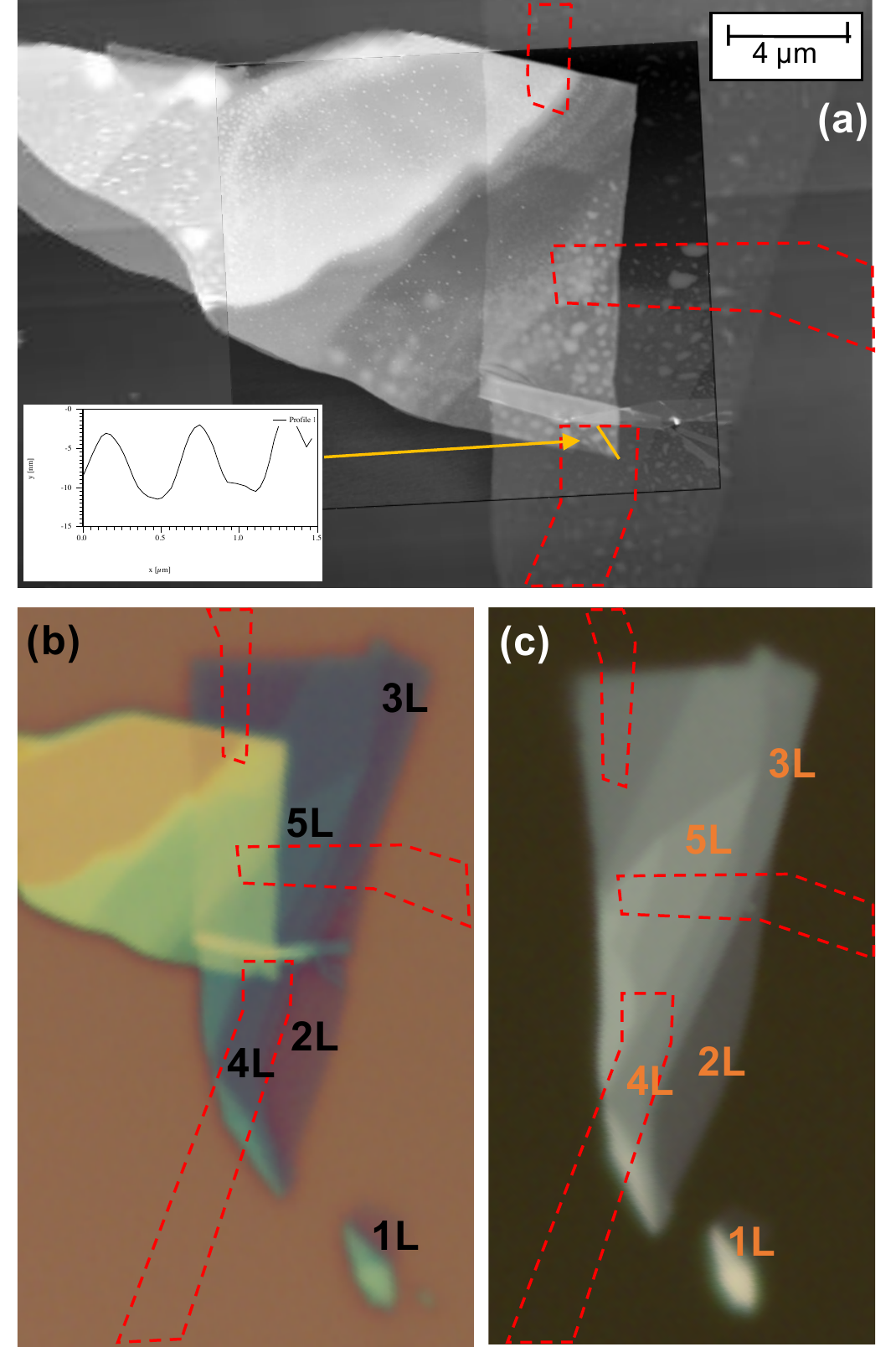}
	\caption[caption]{\textbf{AFM and optical images of the device. a,} AFM imaging of the device discussed in the main text. Position of the tunnel electrodes marked in dashed-red. Inset: cross section of the solid blue line, showing the typical size of the dirt on the device. \textbf{b,} optical image of the two flakes prior to the deposition of the electrodes. Black numbers mark the number of layers observed, from 1 to 5. \textbf{c,} optical image of the \chem{MoS_2} flake on the PDMS prior to the transfer process. Orange numbers mark the number of layers observed, from 1 to 5.  }
	\label{layers}
\end{figure}
The high optical contrast between layers of different thickness of transition metal dichalcogenides (TMDs) allows easy identification of the thickness of the tunnel barrier. Figure \ref{layers} shows the optical image of the barrier on the PDMS immediately after it was exfoliated (panel c) and on top of the \chem{NbSe_2} flake after the transfer procedure (panel b). Both show clearly that the source electrode was deposited above a region consisting of 4 and 5 layer thick \chem{MoS_2}. As a result of exponential dependence of the tunnel current on the barrier thickness, only the 4 layer part of the junction is significant to the measurement. Hence we expect the effective junction area to be 1.6 $\mu m^2$ and the barrier thickness to be between 2.4 nm and 2.6 nm.

Contrary to the optical images, AFM does not provide a reliable measure of height between two different materials and cannot measure the thickness of the barrier. However AFM reveals some structures which are probably due to PDMS residue from the transfer process (panel a). A cross section of some of these features in the area of the junction shows height variation on the scale of 7 nm. The usual cleaning techniques of heat annealing cannot be used here due to the sensitivity of \chem{NbSe_2} to heat. The effect of this structure is most likely to reduce the effective area of the junction to the non-contaminated region. As discussed in the next section, the effective area of the junction is of the same order of magnitude as the observed area, showing the robustness of this method to imperfections.

\section{Detailed derivation of the 2-band model}

The model used to fit the data measured in this work, utilizes the McMillan equations \cite{McMillan1968} similar to the method presented in refs. \cite{Schmidt,Noat2015}. These equations include the pairing amplitudes $\Delta^0_{1,2}$ as fitting parameters. The pairing amplitudes, however, are not fundamental properties - they depend on the BCS coupling within the bands, between the bands, on the interband single electron scattering rates, and on the temperature. $\Delta^0_{1,2}$ are therefore extracted from the fit of Eq. 1 in the main text, but can be calculated from fundamental properties. In this section we outline this calculation, using a model that fully includes all possible effects of two band superconductivity. We then check for consistency between the fit and the calculation.

The model Hamiltonian reads,
\begin{align}
H = H_0 + H_{int} + H_{dis}\, ,
\end{align} 
where the terms, $H_{0}$, $H_{int}$ and $H_{dis}$ describe the band dispersion, Cooper channel interaction and disorder scattering respectively.
We have
\begin{align}\label{H1}
H_0  = &\sum_{k,\sigma}E_k^{1} a^\dag_{k,\sigma} a_{k,\sigma} +\sum_{k,\sigma}E_k^{2}
b^\dag_{k,\sigma} b_{k,\sigma} \\
\begin{split}
H_{int} = & \frac{g_{11}}{2} \sum_{ \mathclap{\substack{k,k' \\ \sigma, \sigma'}} } a_{k\sigma}^\dag a_{-k\sigma'}^\dag a_{-k' \sigma} a_{k' \sigma} +		
\frac{g_{22}}{2} \sum_{ \mathclap{\substack{k,k' \\ \sigma, \sigma'}} } b_{k\sigma}^\dag b_{-k\sigma'}^\dag b_{-k' \sigma} b_{k' \sigma} +\\
&	\frac{g_{12}}{2} \sum_{ \mathclap{\substack{k,k' \\ \sigma, \sigma'}} } a_{k\sigma}^\dag a_{-k\sigma'}^\dag b_{-k' \sigma} b_{k' \sigma} +
\frac{g_{12}}{2} \sum_{ \mathclap{\substack{k,k' \\ \sigma, \sigma'}} } b_{k'\sigma}^\dag b_{-k'\sigma'}^\dag a_{-k' \sigma} a_{k' \sigma}
\end{split}	\\
H_{dis}  = &	\sum_{ \mathclap{\substack{k,k' \\ \sigma}} } ( V_{kk'} a_{k\sigma}^\dag b_{k'\sigma} + V_{k'k} b_{k'\sigma}^\dag a_{k'\sigma}) +
\sum_{ \mathclap{\substack{k,k' \\ \sigma}} } ( \bar{V}^1_{kk'} a_{k\sigma}^\dag a_{k'\sigma} + \bar{V}^2_{k'k} b_{k'\sigma}^\dag b_{k'\sigma})
\end{align}
Here $ a^\dag_{k\sigma}$ and $b^\dag_{k\sigma}$ are the creation operators of the electrons  in Bloch states with momentum $k$ and spin $\sigma$  in  $1$ and $2$ bands respectively.
In what follows, the index $\alpha = 1,2$ labels the bands.
In the Hamiltonian, Eq.~\eqref{H1}, $E_k^{\alpha} $ are the electron dispersion in band $ \alpha $. 
The constants $g_{\alpha\beta}$ are Cooper channel interactions, and $ V_{k'k}, \bar{V}_{k'k}^{\alpha}$ are inter- and intra-band disorder scattering potentials between momenta $k$ and $k'$.
The dimensionless couplings are introduced as
\begin{align}\label{lambda_def}
\lambda_{\alpha\beta} = - g_{\alpha\beta}\sqrt{\nu_{\alpha} \nu_{\beta}}\, .
\end{align}
where $\nu_{\alpha}$ is the normal state density of states in the band $\alpha$. Our sign convention in Eq.~\eqref{lambda_def} is such that positive couplings describe attraction.
The first two terms of $H_{int}$ describe intra-band Cooper pair scattering and the latter two terms describe inter-band scattering. Notice that the intra-band single-electron disorder scattering in $ H_{dis} $ does not affect our results due to the Anderson theorem~\cite{Anderson1959}, it is introduced here for completeness.

We then derive the self consistent equations in Matsubara formalism:
\begin{align}\label{eq:SC_final}
	\begin{split}
		\Delta_1(\epsilon_n) & = \frac{1}{Z_1} \Bigg[ \Delta_1^{0} + \Gamma_{21} \frac{ \Delta_2(\epsilon_n)}{\sqrt{\epsilon_{n}^2+\Delta_2(\epsilon_n)^2} } \Bigg] \\
		\Delta_2(\epsilon_n) & = \frac{1}{Z_2} \Bigg[ \Delta_2^{0} + \Gamma_{12} \frac{ \Delta_1(\epsilon_n)}{\sqrt{\epsilon_{n}^2+\Delta_1(\epsilon_n)^2} } \Bigg]
	\end{split}
\end{align}
where,
\begin{align}\label{eq:delta_ph}
	\begin{split}
		\Delta_1^{0} & = \lambda_{11} 2\pi T \sum_{n'}^{N_m} \frac{\Delta_1(\epsilon_{n'})}{\sqrt{\epsilon_{n'}^2+\Delta_1(\epsilon_n)^2}} + \frac{\lambda_{12}}{\sqrt{\beta} } 2\pi T \sum_{n'}^{N_m} \frac{\Delta_2(\epsilon_{n'})}{\sqrt{\epsilon_{n'}^2+\Delta_2(\epsilon_n)^2}}  \\
		\Delta_2^{0} & =  \lambda_{22} 2\pi T \sum_{n'}^{N_m} \frac{\Delta_2(\epsilon_{n'})}{\sqrt{\epsilon_{n'}^2+\Delta_2(\epsilon_n)^2}} + \lambda_{A2} \sqrt{\beta} 2\pi T \sum_{n'}^{N_m} \frac{\Delta_1(\epsilon_{n'})}{\sqrt{\epsilon_{n'}^2+\Delta_1(\epsilon_n)^2}} 
	\end{split}
\end{align}
and
\begin{align}\label{eq:Z}
	\begin{split}
		Z_1(\epsilon_n) & = 1 + \frac{\Gamma_{21}}{{\sqrt{\epsilon_{n}^2+\Delta_2(\epsilon_n)^2}}}  \\
		Z_2(\epsilon_n) & = 1+ \frac{\Gamma_{12}}{{\sqrt{\epsilon_{n}^2+\Delta_1(\epsilon_n)^2}}} \, .
	\end{split}
\end{align}

We emphasize that this derivation is different from the McMillan derivation by the inclusion of the term $\lambda_{12}$ i.e. Cooper pair tunneling between the bands. This term is irrelevant for the calculation of the proximity effect, but in principle should be present when discussing two band superconductivity.   

We estimate $\Delta^0_{1,2}$ by using the following values: $\Gamma_{12}$ = 1.1, $\Gamma_{21}$ = 0.38 meV (obtained from the fit), and $\lambda_{11}$ = 0.22, $\lambda_{22}$ = 0.13, $\lambda_{12}$ = 0.001, and $\Lambda_D$ = 500 meV. Such a high value for $\Lambda_D$ is not physical. A more realistic value would be $\Lambda_D $ = 60 meV, and  $\lambda_{11}$ = 0.15, $\lambda_{22}$ = 0.01, $\lambda_{12}$ = 0.001. This results in $\Delta^0_1 $ = 1.15 meV, $\Delta^0_2 $ = 0.36 meV. The observed values of $\Delta^0_{1,2}$ are on the high side for weak coupling given $T_C=7.2K$, suggesting that the weak-coupling assumption is only marginally valid.

\section{Intrinsic vs. induced 2$^{nd}$ order parameter}
\begin{figure}[h]
	\centering
	\includegraphics[width=0.5\textwidth]{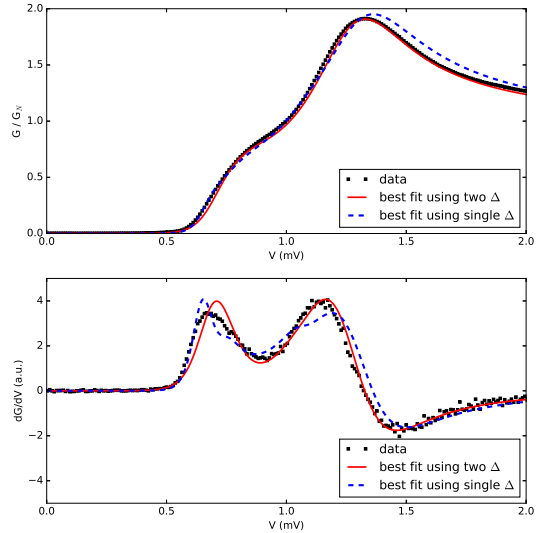}
	\caption{\textbf{Two vs. one intrinsic pairing amplitude} Comparison of fits to the SSM model while allowing for two intrinsic pairing amplitudes (red solid curve)  and when forcing the constraint $\Delta_2 = 0$ (blue dashed curve). The fits are plotted superimposed on \textbf{a.} the $dI/dV$ curve  and \textbf{b.}  the $d^2I/dV^2$ curve.  The fitting parameters values obtained with the two intrinsic  pairing amplitudes is given in the main text. The values obtained with fixed $\Delta_2 = 0 $ are:  $\Delta_1 = 1.26\pm0.01\  \mathrm{meV}$, $\Delta_2 = 0\  \mathrm{meV}$, $\Gamma_{12	} = 0.55 \pm 0.01\ \mathrm{meV}$, $\Gamma_{21}	 = 2.21\pm0.03\ \mathrm{meV}$,  $T = 0.31\pm 0.06 \ \mathrm{K}$, and $\eta = 1:0.05$.  }
	\label{sup_TwoBand}
\end{figure}

Although Noat \textit{et al.}~\cite{Noat2015} report a fit to a single intrinsic pairing amplitude while leaving the other one as induced, the SSM model can intrinsically support two pairing amplitudes. To distinguish between these two scenarios, we fit the measured $dI/dV$ curves in two different modes: (i) without any constraints, thus allowing both pairing amplitudes $\Delta_{1,2}$ as fit parameters, and (ii) while fixing $\Delta_2 = 0$ . The fits obtained are presented in figure \ref{sup_TwoBand} superimposed on the measured data.  While both fits agree reasonably well with the $dI/dV $curve (panel a), it is clear that the 2-order-parameter model fits the data better. This is seen more clearly in the second derivative. Here, the 2-order-parameter model traces the outer peak, while both models fall short of a perfect fit at the inner peak.

\section{2-band fit of tunneling data in magnetic field}

\begin{figure}[h]
	\centering
	\includegraphics[width=1.0\textwidth]{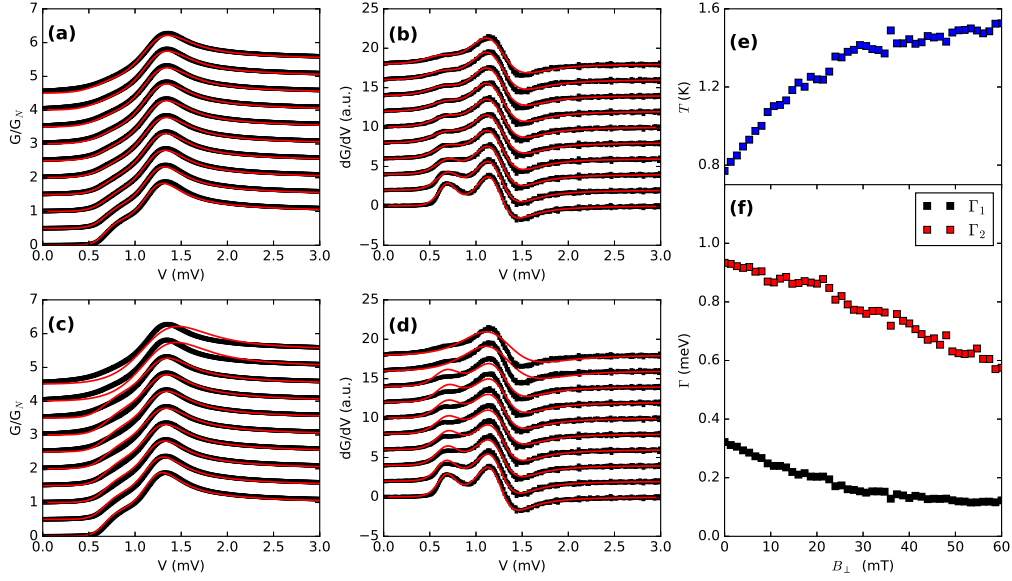}
	\caption{\textbf{SSM model fits at low perpendicular magnetic field } \textbf{a, b} $dI/dV$ and $d^2I/dV^2$ curves fit where $\Delta_{1,2}^0$ are fixed to their zero field values and $\Gamma_{1,2}$ and $T$ are free parameters. \textbf{c, d,} $dI/dV$ and $d^2I/dV^2$ curves fit where $\Gamma_{1,2}$ are fixed to their zero field values and $\Delta_{1,2}^0$ and $T$ are free parameters. \textbf{e,} Field dependence of the effective temperature obtained using the fixed $\Delta_{1,2}^0$ fit. \textbf{f,} Field dependence of the coupling parameters, $\Gamma_{1,2}$, using the fixed $\Delta_{1,2}^0$ fit. 
    }
	\label{sup_FieldFit}
\end{figure}

We fit the $dI/dV$ curves measured with small perpendicular magnetic field using the zero field SSM model (figure \ref{sup_FieldFit}). We begin with testing the more predictable model, where we assume the smaller gap is more fragile to magnetic fields. This model should yield lower $\Delta_2^0$ while keeping the coupling parameters $\Gamma_{12}$ fixed. This, however, clearly fails to fit the data (panels c,d). The fit which does successfully reproduce the experimental curve (panels a,b), involves decreasing the inter-band coupling parameters, $\Gamma_{12}$ and $\Gamma_{21}$ (panel (f)). As these are associated with microscopic scattering processes, we don't expect them to be sensitive to magnetic field, and hence the origin of their suppression remains unclear. The fit also yields an increase in effective temperature (panel (e)). Such temperature broadening of the coherence peaks could be a manifestation of spatially-dependent Abrikosov-Gor'kov (AG) corrections due to the currents around the vortices~\cite{AbrikosovA.A.1961ContributionImpurities,Maki1964TheCurrents}. Quantitative modeling of this effect requires a detailed calculation incorporating the AG corrections into the McMillan model~\cite{Kaiser1970McMillanAlloys}, and calculating the spatially-varying vortex DOS, as has been done (without AG corrections) in Ref.~\cite{Gygi1991Self-consistentSuperconductor} for a single superconducting band.

\section{Estimate of the barrier transparency}

We can estimate the transparency of our tunnel barrier $\mathcal{T}$ from the well-known expression from Sharvin~\cite{Sharvin1965}:

\begin{equation} \label{sharvin-eq}
G_N = \frac{2e^2}{h}\frac{k_F^2 A}{4 \pi}\mathcal{T},
\end{equation}

where $G_N$ is the junction conductance in the normal state, $A$ is the area of the junction, $k_F$ the Fermi momentum and $\mathcal{T}$ the average transmission of each conductance channel. We measure $G_N=$ 7$\mu$S for $A =$ 1.6 $\mu m^2$. $k_F$ in metals is usually $\sim 10^{10}~m^{-1}$ and it is about half this value in \chem{NbSe_2}. Taking the lower value, we get $\mathcal{T}\sim 3\times10^{-8}$.

We can make an independent estimate of $\mathcal{T}$ using the textbook WKB formula for a square barrier of thickness $d$ and height $U$ \cite{Griffiths2005}:

\begin{equation} \label{griffiths}
\mathcal{T} =\exp(-2d\sqrt{2m^*U}/\hbar)
\end{equation}

where $m^*$ is the effective mass of the electron in the barrier, here \chem{MoS_2}.

The gap of few layer \chem{MoS_2} at the $\Gamma$ point in the Brilloiun zone is on the order of 2eV, whereas the effective mass is generally a fraction of 1. Taking $U=$ 1eV, $m^*=m/2$ ($m$ being the bare electron mass), and $d$ in the range 2.4--2.6nm we find $\mathcal{T}\sim3\times 10^{-8}$--$6.5\times 10^{-9}$, consistent with the Sharvin estimate. 

We can make a more rigorous estimate of $U$ (and thus $\mathcal{T}$) by using Brinkman et al.'s result~\cite{Brinkman1970TunnelingBarriers} for the conductance across a trapezoidal barrier with diffuse boundaries, together with measurements of the high bias conductance of our junction:

\begin{equation} \label{brinkman-eq}
\frac{G(V)}{G(0)} = 1-\frac{A_0\Delta\phi}{16\bar{\phi}^{3/2}}eV+\frac{9}{128}\frac{A_0^2}{\bar{\phi}}(eV)^2 
\end{equation}\\

where $V$ is the voltage across the barrier, $\bar{\phi}$ is the mean barrier height, $\Delta\phi$ the barrier height difference on the two sides of the trapezoid, $\mathtt{d}$ the barrier width and $A_0 = 4\sqrt{2m^*}\mathtt{d}/3\hbar$. In these expressions, $\mathtt{d}$ is in units of \AA, while $\bar{\phi}$, $\phi$ and $V$ are in units of volts.

\begin{figure}[h]
	\centering
	\includegraphics[width=0.5\textwidth]{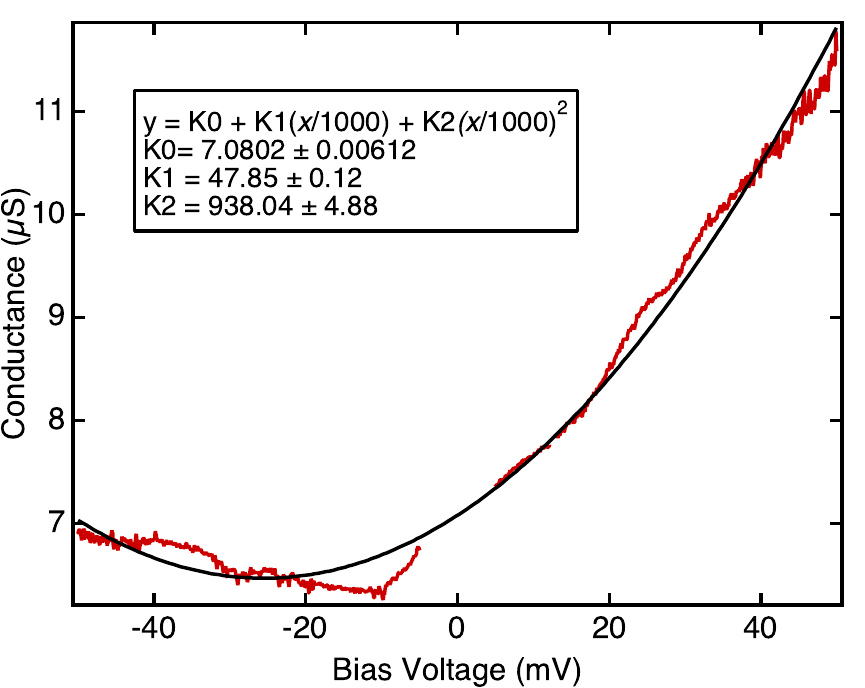}
	\caption[caption]{Conductance as a function of voltage at high biases (red) with a parabolic fit (black). The fit allows us to estimate our barrier height.}
	\label{background}
\end{figure}

Far from the Fermi level, the conductance of our junction indeed rises (Figure~\ref{background}). This rise is not perfectly parabolic and is likely due, in part, to factors other than barrier transparency and asymmetry. Therefore, fitting a parabola to the background, i.e. assuming that the rise is due almost entirely to the barrier, will give us a worst case scenario or minimum possible barrier height.

From the fit to our data to Equation~\ref{brinkman-eq} using $\mathtt{d}$ = 20\AA, we find $\bar{\phi}\sim$0.8V, not so different from what we assumed previously. If we use this, and $d = $ 2.4--2.6nm as before, $\mathcal{T}\sim2\times10^{-7}$--$5\times 10^{-8}$.

Considering all of the above, $\mathcal{T}$ is likely in the $10^{-8}$ range or close to it.

\section{Possible origin of the sub-gap signal}

At zero magnetic field, the sub-gap conductance of the measured tunnel junction at zero bias voltage is highly suppressed, to $\sim$1/500 of the normal state conductance. This residual conductance cannot be accounted for by the thermal broadening of the SSM model, as this gives negligible values at the sub-kelvin temperature range. 

In principle, one or more of the following could be responsible for the sub gap signal: (i) pair tunnelling (due to Andreev reflection) \cite{Blonder1982}; (ii) environment-assisted tunnelling, which can be described by a `Dynes' parameter $\gamma$ in the BCS density of states (an imaginary part in the energy)\cite{Pekola2010Environment-AssistedStates,Marco2013}; (iii) a finite quasiparticle lifetime, due e.g. to strong electron-phonon coupling, which is also described by a Dynes parameter \cite{Dynes1978}; and (iv) a parallel resistance in the junction. 

\begin{figure}[h]
	\centering
	\includegraphics[width=0.5\textwidth]{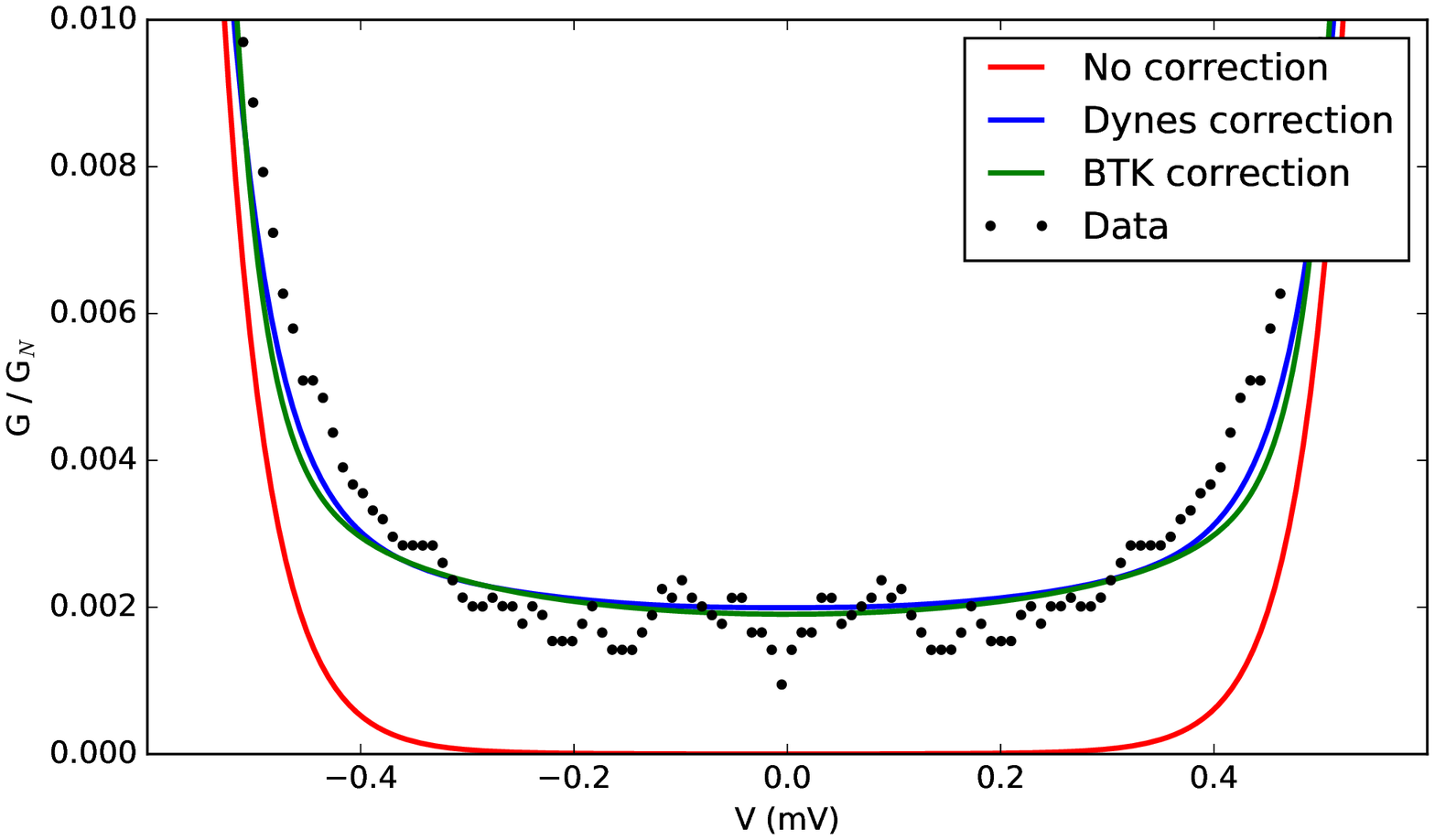}
	\caption[caption]{\textbf{Zero field sub gap conductance } The zero field sub gap conductance data (black dots), compared with the plain SSM model(red curve), with the SSM model corrected using Dynes parameter (blue curve, $\Gamma_\textrm{Dynes} = 0.002$ ), and with the SSM model corrected using the BTK model (green curve, Z = 16). }
	\label{sup_AD}
\end{figure}

In Figure~\ref{sup_AD}, we show the conductance of the junction at energies below the gap and compare it to two theoretical calculations. A fit to the data using the the SSM model with the inclusion of the Dynes terms produces the curve shown in green, which shows good agreement. 

If, on the other hand, we assume that the conductance at $V_{DC}=0$ is due entirely to Andreev processes, using the Blonder, Tinkham and Klapwijk (BTK) model~\cite{Blonder1982} -- using a single band BCS density of states with $\Delta = 0.7$ meV, the apparent size of the gap, and normalised to the measured $G_N$ -- we obtain a curve which also fits the data in the low bias region. However, we find $Z = 16$ for the dimensionless barrier strength, which corresponds to a transparency $\mathcal{T} = 1/(1+Z^2) = 4\times 10^{-3}$, much greater than the value obtained in the previous section and thus not plausible. 

In addition, we note that in Al/\chem{Al_2O_3}/Al NIS junctions (200nm $\times$ 200nm, 5--10k$\Omega$) measured in the same dilution refrigerator with the same measurement setup, $G(0)/G_N$ is typically 1/200. According to Ref. \cite{Pekola2010Environment-AssistedStates}, for a given environment, the environmental contribution to the subgap conductance should be suppressed if either $\Delta$ or $C$, the capacitance of the junction, increases. As  both of $\Delta$ and $C$ of the \chem{NbSe_2} device are higher than the Al device, we expect further suppression of $G_0 R_N$, and this is indeed what we see.

All of the above would seem to suggest that the subgap conductance is not limited by Andreev processes but rather by the environment. We cannot, however, rule out the contribution of finite quasiparticle lifetimes or a resistance parallel to the junction.

\section{Possible origin of the effective temperature}

As mentioned in the main text and as can be seen in Figure \ref{sup_Teff} below, the $dI/dV$ curve of our device agrees very well with the SSM model, with an effective temperature of 500mK --- significantly higher than the known base electron temperature of the dilution refrigerator used, which is $\sim$100mK. This could be due to some combination of: (i) heating from the measurement; (ii) a slight $k$-space anisotropy in $\Delta$; and (iii) defects in the \chem{NbSe_2} such as Se vacancies leading to spatially inhomogeneous $\Gamma$'s and $\Delta$'s. We explore each of these explanations in turn.

\begin{figure}[h]
	\centering
	\includegraphics[width=0.5\textwidth]{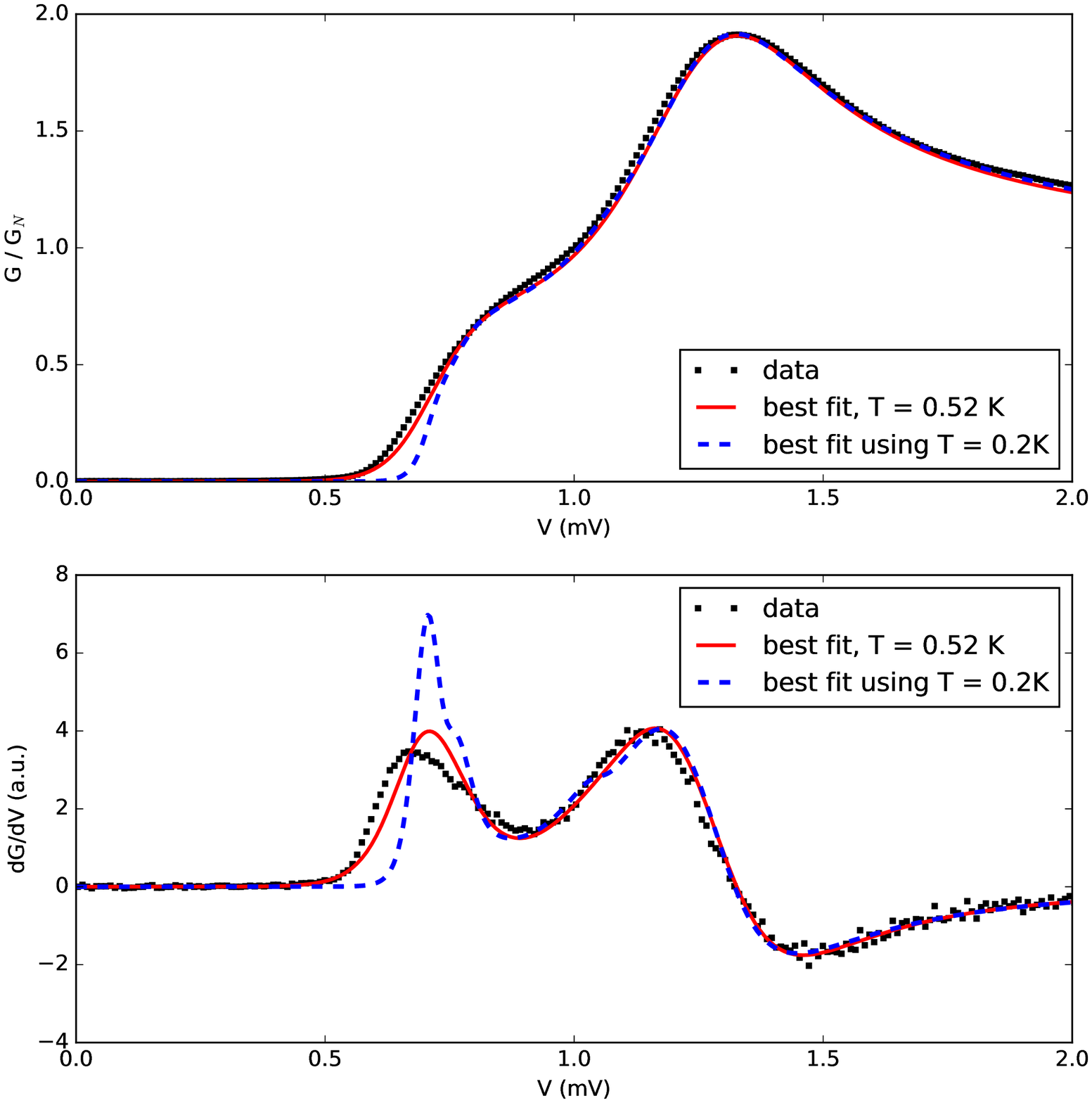}
	\caption[caption]{\textbf{Effective temperature vs. maximal actual temperature} Comparison of fits to the SSM model while treating the temperature as an effective free parameter (red solid curve)  and when forcing the constraint T = 200mK (blue dashed curve). The fits are plotted superimposed on \textbf{a.} the $dI/dV$ curve  and \textbf{b.}  the $d^2I/dV^2$ curve.  The fitting parameters values obtained using the temperature as a free parameter are given in the main text. The values obtained with fixed T = 200 mK are:  $\Delta_1 = 1.24\pm0.01\  \mathrm{meV}$, $\Delta_2 = 0.38 \pm 0.02  \mathrm{meV}$, $\Gamma_{12	} = 0.46 \pm 0.01\ \mathrm{meV}$, $\Gamma_{21}	 = 1.19\pm0.06\ \mathrm{meV}$,  $T = 0.2\ \mathrm{K}$, and $\eta = 1:0.07$. 
	}
	\label{sup_Teff}
\end{figure}

Equating the heat produced by the junction ($IV$) with the heat carried away by the leads ($\kappa\Delta T d/N_\square$) and using the Wiedemann-Franz law ($\kappa = \sigma L T$), we obtain

\begin{equation}\label{temperature-eq2}
\Delta T = \frac{IVR}{LT}.
\end{equation}

Here $I$ is the current and $V$ the voltage across the junction. $\kappa$ is the thermal conductivity, $d$ the thickness and $R$ the resistance of the leads; $N_\square$ the number of squares in the leads and $\Delta T$ the temperature difference across the leads. $L$ is the Lorenz number. 

As the discrepancy between the constrained and unconstrained fits in Figure~\ref{sup_Teff} are most evident well below 1mV, we take $V$= 1mV, $I$ = 3nA and $T=100mK$. $R$ measured at 4K is 20$\Omega\pm$5$\Omega$ and can only decrease at the base temperature of the refrigerator. This yields $\Delta T\sim 25mK$, which is too small to explain the observed effective temperature.

It would therefore seem that gap anisotropy (in $k$-space) or inhomogeneity (in real space) is responsible for the observed effective temperature.

If we assume that $k$-space gap anisotropy is the only mechanism responsible for the effective temperature, we can put an upper bound on the gap anisotropy of \chem{NbSe_2} in the $ab$ plane, by assuming that $\Delta_\textrm{max} -\Delta_\textrm{min} \sim$ the FWHM of the derivative of the Fermi-Dirac function.
\begin{equation}
\alpha =  \frac{
\Delta_\textrm{max} -\Delta_\textrm{min} } {\Delta}  \sim \frac{3.5k_BT_\mathrm{eff}}{\Delta} \sim 0.1.
\end{equation}
We note that this level of anisotropy is indistinguishable from perfect isotropy in its effect on the slope of the zero bias conductance as a function of field (cf. Fig. 2b of main text).

Finally, and more speculatively, we note that an effect such as a $k$-space dependence of the inter-band couplings could also be responsible for $T_{eff}$.

\end{document}